\documentclass[12pt]{article}
\usepackage{epsfig}
\usepackage{amssymb,amsmath}

\def\beq{\begin{equation}}
\def\eeq#1{\label{#1}\end{equation}}
\def\eeqn{\end{equation}}

\def\beqa{\begin{eqnarray}}
\def\eeqa#1{\label{#1}\end{eqnarray}}
\def\eeqan{\end{eqnarray}}

\let\bar=\overbar

\def\etal{{\it et al.}}

\def\Dslash{\not{\hbox{\kern-4pt $D$}}}
\def\dslash{\not{\hbox{\kern-2pt $\del$}}}

\def\msb{{\bar{\ssstyle M \kern -1pt S}}}

\def\Title#1{\begin{center} {\Large {\bf #1} } \end{center}}

\newcommand{\CP}{\ensuremath{\mathit{CP}}}
\newcommand{\ppbar}{\ensuremath{\bar{p}p}}                                                                                       %
\newcommand{\Bd}{\ensuremath{B^{0}}}                                                                                                   %
\newcommand{\Bu}{\ensuremath{B^{+}}}                                                                                                   %
\newcommand{\Bs}{\ensuremath{B_{s}^{0}}}                                                                                               %
 \newcommand{\Lb}{\ensuremath{\Lambda^0_b}}                                                                                     %
\newcommand{\BdKpi}{\ensuremath{\Bd\rightarrow K^{+}\pi^{-}}}                                                           %
\newcommand{\BsKpi}{\ensuremath{\Bs\rightarrow K^{-}\pi^{+}}}                                                              %
                       
\newcommand{\Lbppi}{\ensuremath{\Lb \rightarrow p\pi^{-}}}   
\newcommand{\LbpK}{\ensuremath{\Lb \rightarrow p K^{-}}}  

\newcommand{\lumifb}{ fb$^{-1}$}

\newcommand{\acpbdkpi}{\ensuremath{A_{\CP}(\BdKpi)}}
\newcommand{\acpbskpi}{\ensuremath{A_{\CP}(\BsKpi)}}
\newcommand{\acpLbppi}{\ensuremath{A_{\CP}(\Lbppi)}}
\newcommand{\acpLbpK}{\ensuremath{A_{\CP}(\LbpK)}}

\newcommand{\Bsphiphi}{\ensuremath{B_s^0 \to \phi \phi}}
\newcommand{\tp}{\ensuremath{\mathrm{TP}}}

\newcommand{\atp}{\ensuremath{\mathcal{A}_{\mathrm{TP}}}}

\begin{document}

\noindent
\begin{flushright}
{\small
Proceedings of CKM 2012, the $7^{\rm th}$ International Workshop on the
CKM Unitarity Triangle, University of Cincinnati, 
USA, 28 September - 2 October 2012.
}
\end{flushright}
\bigskip

\Title{TeVatron Direct \CP\ Violation Results} 

\bigskip\bigskip


\begin{raggedright}  

{\it Michael J. Morello\index{Morello, M.J.}\\
Scuola Normale Superiore and INFN Pisa\\
I-56127 Pisa, ITALY}
\bigskip\bigskip
\end{raggedright}

\begin{abstract}
I report some recent results on direct \CP\ violation measurements in hadronic decays collected by
the upgraded Collider Detector (CDF II) at the Fermilab Tevatron: 
 \CP-violating asymmetries in the two-body non-leptonic charmless decays of $b$-hadrons,  
the first reconstruction in hadron collisions of the suppressed decays
$B^- \to D(\to K^+\pi^-)K^-$ and $B^- \to D(\to K^+\pi^-)\pi^-$,
and the measurement of TP asymmetries in the \Bsphiphi\ decays.
\end{abstract}

\section{Introduction}

Non-invariance of the fundamental interactions under the combined
symmetry transformation of charge conjugation and parity inversion (\CP\ violation) 
is an established experimental fact. The vast majority of experimental data are well described by the standard model (SM), 
and have supported the success of the Cabibbo-Kobayashi-Maskawa (CKM) 
theory of quark-flavor dynamics. However, additional sources of \CP\ violation are required to explain
the matter--antimatter asymmetry of the Universe in standard big-bang
cosmology. The  heavy flavour sectors have not yet been fully covered
by experiments so far, thus the presence of a new 
source of \CP\ violation can not be excluded. An unexpected hint of \CP\ violation would have profound consequences on our understanding 
of fundamental interactions.

The CDF II experiment at the Tevatron $p\bar{p}$ collider established that extensive and detailed exploration of the $b$--quark 
dynamics is possible in hadron collisions, with results competitive and supplementary to those from $e^+e^-$ colliders. 
This has provided a rich and highly rewarding physics program 
and, still more important, a tremendous legacy for currently operating and future experiments, such as LHCb.

\section{Two-body non-leptonic charmless B decays}

In recent times, the pattern of direct \CP\ violation in
charmless mesonic decays of $B$ mesons has shown some 
unanticipated discrepancies from expectations. 
Under standard assumptions of isospin symmetry and smallness of contributions 
from higher-order processes, similar \CP\ asymmetries are predicted for \BdKpi\ and $\Bu\to K^+\pi^0$ decays~\cite{Keum:2002vi,Beneke:2003zv}. 
However, experimental data show a significant discrepancy~\cite{acp_bfactories}, 
which has prompted intense experimental and theoretical activity. Several simple extensions of the SM
could accommodate the discrepancy~\cite{newphysics}, but  
uncertainty on the contribution of higher-order SM amplitudes 
has prevented a firm conclusion~\cite{allstandard,Lipkin:2011hh}. 
Measurements of direct \CP\ violation in  \BsKpi\ 
decays have been proposed as a nearly model-independent test
for the presence of non-SM physics~\cite{Gronau:2000md,Lipkin:2005pb}. 
The relationships between charged-current quark 
couplings in the SM predict a well-defined hierarchy between direct 
\CP\ violation in \BdKpi\ and \BsKpi\ decays, yielding a significant asymmetry for the latter, of 
about 30\%. This large effect allows easier experimental 
investigation and any discrepancy may indicate contributions from non-SM amplitudes.
Supplementary information could come from \CP\ violation 
in bottom baryons. Interest in charmless $b$--baryon decays is prompted by branching fractions recently 
observed being larger than expected \cite{Mohanta:2000nk, Lu:2009cm, Aaltonen:2008hg}. 
Asymmetries up to about 10\% are predicted for \LbpK\ and \Lbppi\ decays 
in the SM~\cite{Lu:2009cm,Mohanta:2000za}, and are accessible with current available samples.
High precision measurements of the violation of \CP\ symmetry in charmless modes remains, therefore, a very interesting 
subject of study and may provide useful information to our comprehension of this discrepancy.
\begin{figure}[htb]
\begin{center}
\epsfig{file=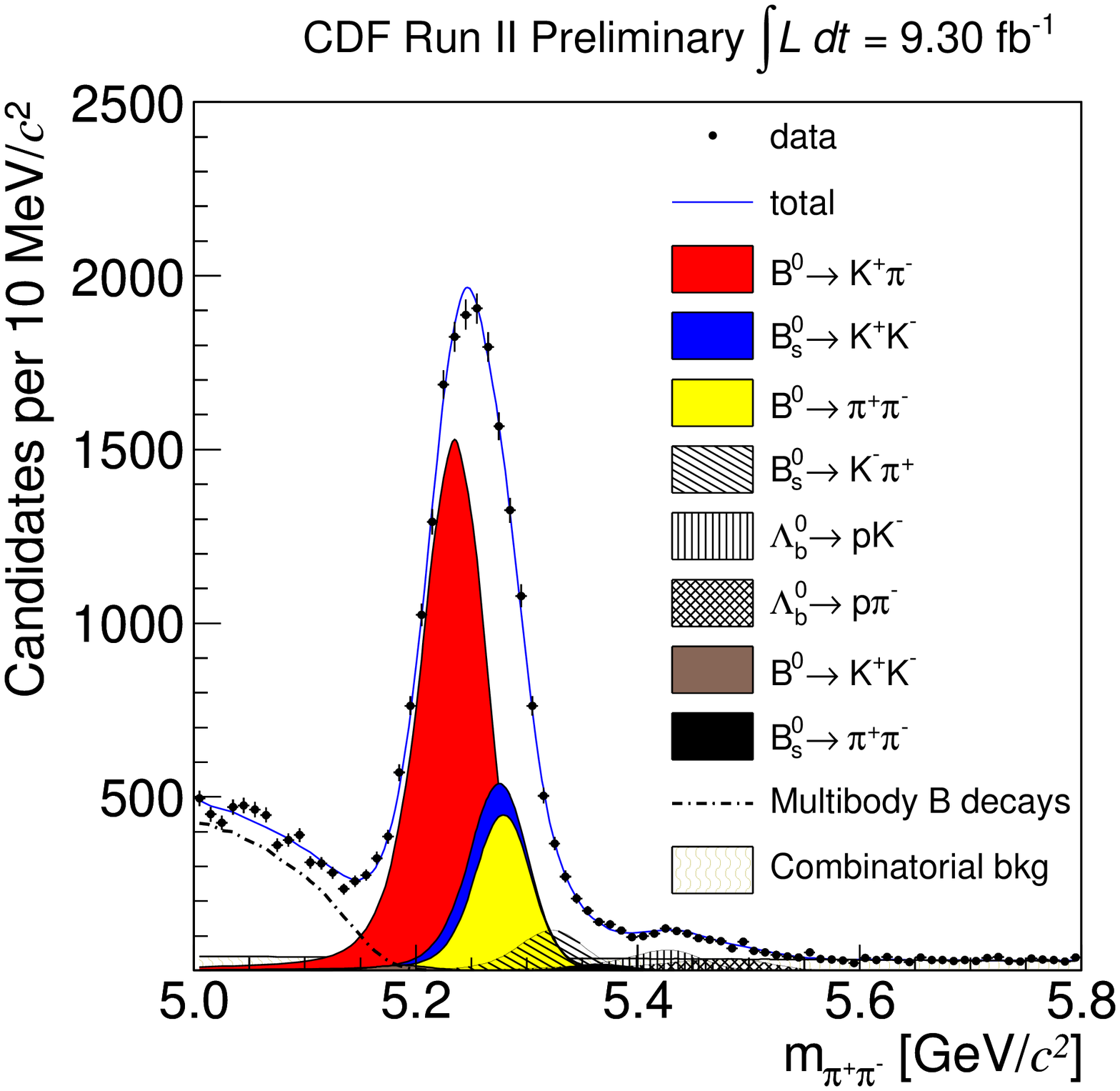,height=2.8in}
\epsfig{file=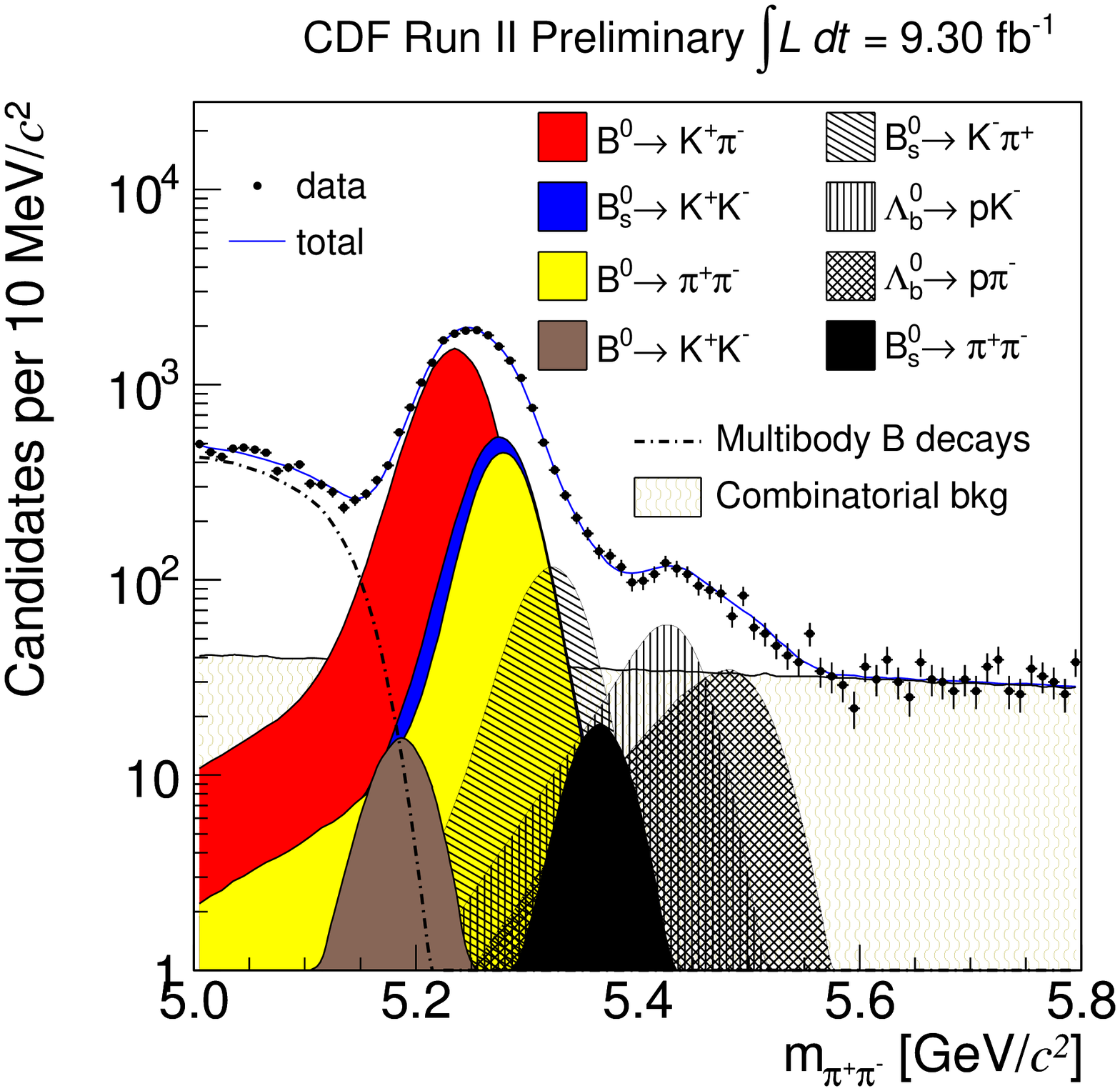,height=2.8in}
\caption{\label{fig:projections} Mass distribution of reconstructed candidates, $m_{\pi\pi}$.  
The charged pion mass is assigned to both tracks. 
The total projection and projections of each signal and background component of the likelihood fit are overlaid on the data distribution.
Signals and multi-body $B$ background components are shown stacked on the combinatorial background component.
Linear scale (left panel) and logarithmic scale (right panel).
}
\label{fig1} 
\end{center}
\end{figure}

We report the measurements of direct \CP\ violation in decays of bottom mesons and bottom baryons, performed in 9.3\lumifb\ of \ppbar\ 
collisions at $\sqrt{s} = 1.96$ TeV, collected  by CDF~II at the Fermilab Tevatron.
An extended unbinned likelihood fit~\cite{cdf10726,Aaltonen:2011jv}, incorporating kinematic (invariant mass and momenta) 
and particle identification (dE/dx) information, is used to determine the
fraction of each individual 
mode in the sample. The fit projection on the invariant $\pi\pi$-mass is reported in fig.~\ref{fig1}.
We measure $\acpbdkpi = -0.083 \pm 0.013 {\rm (stat)} \pm 0.003 {\rm (syst)}$~\cite{cdf10726} 
with a significance more than $5\sigma$. The uncertainty 
of the observed asymmetry is consistent and of comparable accuracy with current results from asymmetric $e^+e^-$ 
colliders~\cite{acp_bfactories} and LHCb~\cite{Aaij:2012qe}.
We also measure  $\acpbskpi = +0.22 \pm 0.07 {\rm (stat)}  \pm 0.02 {\rm (syst)}$~\cite{cdf10726} with a significance of 2.9$\sigma$. 
This result confirms  the LHCb evidence~\cite{Aaij:2012qe} with the same level of resolution. 
The averaged value between this result and LHCb measurement is equal to $\acpbskpi_{\rm mean}=+0.24 \pm 0.05$
which has a significance of 4.8$\sigma$. This represents a strong evidence of \CP\ violation in the $B^0_s$ meson system.
The observed asymmetries  $\acpLbpK=-0.09 \pm 0.08 {\rm (stat)}  \pm 0.04 {\rm (syst)}$~\cite{cdf10726}  and 
$\acpLbppi = +0.07 \pm 0.07 {\rm (stat)} \pm 0.03 {\rm (syst)}$~\cite{cdf10726} are consistent with zero. 
The current experimental precision allows for the first time to exclude large \CP\ violation effects in these decays, however 
 it is not yet sufficient for a conclusive discrimination between the standard 
model prediction (8\%) and much suppressed values ($\approx 0.3\%$) expected in 
R--parity violating supersymmetric scenarios~\cite{Mohanta:2000za}. 
The observed asymmetries are consistent with the previous results from CDF~\cite{Aaltonen:2011qt} and supersede them.

\section{Angle \boldmath{$\gamma$} from \boldmath{$B^{-} \to DK^{-}$}}

Conventionally, CP violating observables are written in terms of the angles $\alpha$, $\beta$ and $\gamma$ of 
the Unitarity Triangle, obtained from one of the unitarity conditions of the CKM matrix. 
While the resolution on $\alpha$ and $\beta$ reached a good level of precision, the measurement of $\gamma$ is still limited 
by the smallness of the branching ratios involved in the processes. 
Among the various methods for the $\gamma$ measurement, those which make use of the tree-level $B^- \to D^0 K^-$ 
decays have the smallest theoretical uncertainties. 
In fact $\gamma$ appears as the relative 
weak phase between two amplitudes, the favored $b \to c \bar{u} s$ transition of the $B^- \to D^0 K^-$, whose amplitude is 
proportional to $V_{cb} V_{us}$, and the color-suppressed $b \to u \bar{c} s$ transition of the $B^- \to \overline{D}^0 K^-$, 
whose amplitude is proportional to $V_{ub} V_{cs}$.  
The interference between $D^0$ and $\overline{D}^0$, decaying into the same final state, leads to measurable \CP-violating 
effects, from which $\gamma$ can be extracted. 
The effects can be enhanced by choosing interfering amplitudes that
are of the same order of magnitude.
All methods require no tagging or time-dependent measurements, and many of them only involve charged particles in 
the final state. 
%

In a data sample of about 7~\lumifb\ 
we report the first reconstruction in hadron collisions of the suppressed decays 
$B^- \to D(\to K^+ \pi^-)K^-$ and $B^- \to D(\to K^+ \pi^-)\pi^-$, which are the main ingredient of 
the ADS method~\cite{ref:ads1}.
Also in this case an extended unbinned likelihood fit,
incorporating kinematic (invariant mass) 
and particle identification (dE/dx) information, is used to determine the
fraction of each individual modes. 
CDF measures the following asymmetries:
$A_{ADS}(K)  =   -0.82 \pm 0.44\mbox{(stat)} \pm 0.09\mbox{(syst)}$ and 
$A_{ADS}(\pi) =   0.13 \pm 0.25\mbox{(stat)} \pm 0.02\mbox{(syst)}$~\cite{cdf_ads},
and for the ratios of doubly Cabibbo suppressed mode to flavor eigenstate CDF finds 
$R_{ADS}(K)  =  [22.0 \pm 8.6\mbox{(stat)} \pm 2.6\mbox{(syst)}] \times 10^{-3}$ and 
$R_{ADS}(\pi)  =  [2.8 \pm 0.7\mbox{(stat)} \pm 0.4\mbox{(syst)}] \times 10^{-3}$~\cite{cdf_ads}.
The results are in agreement with existing measurements 
performed at $\Upsilon$(4S) resonance~\cite{ads_bfactories} and very recently at LHCb~\cite{Aaij:2012kz}.

\section{\boldmath{$B^0_s\to \phi\phi$} }

Triple product (TP) asymmetries are odd under time-reversal,  
and can be generated either by final state interactions or \CP\ violation.  
In flavor untagged samples, where the initial $B$ flavor 
is not identified, TP asymmetries can be shown to
signify genuine \CP\ violation~\cite{TPtruefake}.  In this respect they are very sensitive
to the presence of new physics in the decay since they do not require a 
strong-phase difference between new and SM amplitudes, as opposed to
direct \CP\ asymmetries~\cite{TPtheory}.
The TP asymmetry is defined as
$
\atp=\frac{\Gamma(\tp>0)-\Gamma(\tp<0)}
{\Gamma(\tp>0)+\Gamma(\tp<0)},
$
where $\Gamma$ is the decay width for the given process.
In \Bsphiphi\ decays two TP asymmetries can be studied, corresponding to the two interference terms
between amplitudes with different \CP. These asymmetries are predicted to vanish in the SM, and 
an observation of a non--zero asymmetry would be an unambiguous
sign of NP~\cite{TPtheory}.

We report the first measurement  of TP asymmetries in the \Bsphiphi\ decays 
reconstructed at CDF,  using a data sample of  2.9~fb$^{-1}$ of integrated luminosity and
we measure $A_u =-0.007 \pm 0.064 \pm 0.018$ and  $A_v  = -0.120 \pm 0.064 \pm 0.016$~\cite{Aaltonen:2011rs} 
in agreement  with recent world's best measurement from LHCb~\cite{Aaij:2012ud}.

\section{Conclusion}

CDF experiment at the Tevatron keeps providing excellent results in the exploration of
Heavy Flavor Physics,  owing to \CP-symmetric initial states in $\sqrt{s}=1.96$ TeV $p\bar{p}$ collisions, 
large event samples collected by well-understood detector, and mature analysis techniques.
In summary, this short write-up reports on the measurements of \CP-violating asymmetries in the 
two-body non-leptonic charmless decays of $b$-hadrons,  
on the first reconstruction in hadron collisions of the suppressed decays $B^- \to D(\to K^+\pi^-)K^-$ and $B^- \to D(\to K^+\pi^-)\pi^-$, 
and on the measurement of TP asymmetries in the \Bsphiphi\ decays.

\end{document}